\documentclass[useAMS]{mn2e_modmargin}
\usepackage{amsmath}
\usepackage{url}
\usepackage{amsfonts}
\usepackage{amsbsy}
\usepackage[dvips]{graphics}
\usepackage{subfigure}
\usepackage{verbatim}
\usepackage{amssymb}
\usepackage{amsbsy}

\renewcommand{\d}{\ensuremath{\partial}}

\title[The ballistic transport instability]{The ballistic transport
instability in Saturn's rings III:
numerical simulations}
\author[Latter, Ogilvie \& Chupeau]{Henrik N. Latter$^{1}$\thanks{E-mail:
    hl278@cam.ac.uk},
   Gordon I. Ogilvie$^{1}$,
   Marie Chupeau$^{1,2}$ \\
$^{1}$ DAMTP, University of Cambridge, CMS, Wilberforce Road,
Cambridge CB3 0WA, UK\\
$^{2}$ LPTMC, Universit\'e Pierre-et-Marie-Curie, Tour 24, 4, Place
Jussieu, 75252, Paris Cedex 05, France }
\date{}

\begin{document}

\maketitle

\begin{abstract}

Saturn's inner B-ring and its C-ring support wavetrains of contrasting
 amplitudes but with similar length scales,
100---1000 km. 
In addition, the inner B-ring is punctuated by two intriguing
`flat' regions between radii 93,000 km and 98,000 km in which the waves die out, whereas the 
C-ring waves coexist with a forest of
plateaus, narrow ringlets, and gaps. In both regions the waves are
probably generated by a large-scale linear instability whose origin lies in
the meteoritic bombardment of the rings: the ballistic transport
instability. 
In this paper, the third in a
series, we numerically simulate the long-term nonlinear evolution of this instability
in a convenient local model. 
Our C-ring simulations confirm that
the unstable system forms low-amplitude wavetrains possessing
a preferred band of wavelengths.
B-ring simulations, on the other hand, exhibit localised nonlinear
wave `packets' separated by linearly stable flat zones. Wave packets
travel slowly while spreading in time, a result
that suggests the observed flat regions in Saturn's B-ring are shrinking.
Finally, we present exploratory runs of the inner B-ring edge which
reproduce earlier numerical results: ballistic transport can maintain
the sharpness of a spreading edge while building a `ramp' structure at
its base. Moreover, the ballistic transport instability can afflict the ramp
region, but only in low-viscosity runs.
\end{abstract}


\section{Introduction}

Planetary rings of low and intermediate optical depth are vulnerable
to the ballistic transport instability (BTI) which
issues from the continual bombardment of ring particles by
hypervelocity micrometeoroids. Ejecta released via these impacts
reaccrete on to the ring at various radii, and thus redistribute
mass and angular momentum (Durisen 1984, Ip 1984, Lissauer 1984).
A small positive perturbation in surface density will change the
ring's local transport
properties, and 
if the overdense region releases less material than it can absorb
relatively, then it will grow and an instability results
(Durisen 1995, Latter et al.~2012,
hereafter Paper 1). The BTI 
favours long lengthscales
$l_\text{th}\sim 10-10^3$ km and long timescales $t_e \sim 10^5-10^7$ yr. Thus the 100-km waves in
the inner B-ring (between radii 93,000 and 98,000 km) and the 1000-km
 undulations in the C-ring (between 77,000 and 86,000 km) are
possible manifestations of its nonlinear development (Figs 13.17 and
13.13 in Colwell et al.~2009, Charnoz et
al.~2009 ).

This is the third paper in a series exploring the dynamics of the BTI
and its generation of wave-like structure. The first paper presented a
convenient local model with which to study the problem and worked
through the BTI's linear theory (Paper 1). The second paper
established semi-analytically that the instability could sustain families
of nonlinear travelling wavetrains (Latter et al.~2013, herafter Paper
2). For C-ring parameters, these waves
saturate at low amplitude.
For B-ring parameters, the ring exhibits bistability, with the system
 falling into one of two linearly stable states: the background
homogeneous state (a `flat zone') or a large-amplitude wave state (a
`wave zone'). Both results are consistent with Cassini data
and strengthen the connection between the observed wave features and the
BTI. 

In this paper, our earlier results are verified and extended with
time-dependent simulations. 
Our numerical algorithm exploits the convolution form of the
ballistic transport integrals and, as a result, 
can easily evolve the system on extremely long timescales $\sim 1000\, t_e$ and
lengthscales $\sim 100\, l_\text{th}$. We subsequently
 calculate structure
formation in three contexts: (a) low-optical depth models of the
C-ring, (b) bistable models of the B-ring, and (c) 
the spreading and structure of a sharp edge. 

The low-optical depth simulations 
fulfil most of the expectations of Paper 2. After an
initial period of wave competition, the system settles upon a linearly stable low-amplitude
wavetrain that fills the
domain. Our simulations, however, possess translational symmetry, 
a special constraint not shared by the C-ring.
In order to eliminate its effects, we perform additional runs with `buffered'
boundaries that work similarly to out-going wave conditions. Low-amplitude
wavetrains dominate these simulations as well, but their dynamics is more
complicated; in particular, wave activity can propagate out of the
ring entirely leaving behind a state of very low amplitude and 
long wavelength. 
Concurrently waves suffer strong time-dependent inhomogeneities in
their amplitude and phase. Both sets of simulations are
consistent with Cassini observations of the very long C-ring
undulations, and reinforce the attribution of these features to the
BTI.  

Our second group of simulations probes the dynamics of hysteresis in
models of the B-ring. As the system supports both a stable homogeneous
state and a wave state, we set up an initial condition in which these
two states occupy different regions within the computational
domain. We find that the `wave zones' behave like wave packets, 
moving at the group velocity of their constituent waves. In
addition, the wave packets spread, a nonlinear effect
due to the variation of the group velocity through the packet. 
This behaviour agrees qualitatively with the observations, though
interesting discrepancies exist which we discuss. 

Finally, exploratory runs of a spreading ring edge are presented.
Starting from a step function in optical depth between very thin and
thick regions, we confirm the finding of Durisen
et al.~(1992, hereafter D92), that ballistic transport maintains the
sharpness of an edge as it spreads, while building at its foot a `ramp' (i.e.\ a
region with a shallow optical depth gradient). Lower viscosity runs,
however, indicate that this ramp is unstable to the BTI, with growing
modes reaching large amplitudes. As such structures are absent
in the Cassini observations, it may be possible to roughly constrain ring
properties from this result.

The organisation of the paper is as follows. In Section 2 we summarise
the mathematical details of our physical model as well as the
numerical method with which we solve it. Section 3
presents simulations that approximate the C-ring, Section 4
deals with the B-ring simulations, while Section 5 contains our results on
spreading ring edges. We draw our conclusions and point to
future work in Section 6.

\section{Governing equations and numerical set-up}
\subsection{Mathematical formalism}

Our simulations are undertaken in the shearing box, which is a local
model of a planetary ring that ignores curvature effects and global 
gradients in ring properties. The box is centred at a fixed radius $r_0$,
with $x$ denoting the local radial co-ordinate. Its radial size is
$L$, and we must supply (potentially unrealistic)
boundary conditions.  

Following Paper 1, 
the time evolution of the dynamical optical depth $\tau$ is given by
the following integro-differential equation:
\begin{align} \label{GovEq}
\d_t \tau = \mathcal{I}-\mathcal{J} +
\tfrac{1}{2}\d_x\left(\mathcal{K}+\mathcal{L}\right) + \d_x(\mu\d_x \tau),
\end{align} 
where $\mu$ is a measure of the relative strength of viscous
over ballistic transport. In contrast to Papers 1 and 2, we
permit $\mu$ to vary with space in some runs. The integral operators
$\mathcal{I}$ and $\mathcal{J}$ describe the direct transfer of
mass by ballistic processes, while $\mathcal{K}$ and $\mathcal{L}$
describe the transfer of angular momentum. In Equation
\eqref{GovEq}, the units of time and space have been chosen so that the
characteristic ballistic throw length $l_\text{th}$ and the
characteristic ballistic erosion time $t_e$ are
1 (see Papers 1 or 2 for their exact definitions).

The integral operators in the governing equation may be compactly expressed using
convolutions
\begin{align} \label{coll1}
&\mathcal{I} = P\cdot (R*f), &\mathcal{J}= R\cdot (P*\tilde f), \\
&\mathcal{K} = P\cdot (R*g), & \mathcal{L} = R\cdot(P*\tilde g).\label{coll2}
\end{align}
Appearing in these expressions are the three key functions describing
ballistic transport: the rate of
ejecta emission per unit time and area $R(\tau)$, the probability of
mass absorption from incoming ejecta $P(\tau)$, and the ejecta distribution
function $f(x)$, defined so that $f(x)dx$ is the proportion of
material thrown distances between $x$ and $x+dx$. 
The function $g$ is defined by $g(x)=x f(x)$,
 and the tilde denotes a reflection, so that $f(x)=\tilde f(-x)$ and
$g(x)=\tilde
g(-x)$. Note that more generally, $P$
is a function of both $\tau$ at the absorbing radius and
$\tau$ at
the emitting radius.

As in Papers 1 and 2, the functional
forms for ejecta emission $R$ and absorption $P$ are:
\begin{align}
&P(\tau)= 1-\text{exp}(-\tau/\tau_p), \label{P}\\
&R(\tau) = 0.933\left[ 1 + \left(\frac{\tau}{\tau_\mathrm{s}}-1\right)
\exp(-\tau/\tau_\mathrm{s})\right], \label{R}
\end{align}
the latter taken from Cuzzi and Durisen (1990).
We fix the parameters so that the reference optical
depths are $\tau_p=0.5$ and
$\tau_s=0.28$. The throw distribution calculated by Cuzzi \& Durisen (1990) is
approximated by an off-centred Gaussian profile,
\begin{align}
f(\xi) = \frac{1}{\sqrt{2\pi d^2}}\,\text{exp}\left[-(\xi-\xi_0)^2/(2d^2)
\right], \label{f}
\end{align}
in which we set the offset to $\xi_0=0.5$ and the standard
deviation to $d=0.6$. 

\subsection{Numerical approach}

We employ a third-order Runge-Kutta time stepper to evolve
Eq.~\eqref{GovEq} while computing the spatial derivatives and
integrals with a Fourier pseudo-spectral method. Consequently, the
radial domain is partitioned into $N$ nodes, each equally spaced by
$\Delta x$. The $x$-derivatives of $\tau$ are calculated in Fourier
space, using a FFT routine, and the integrals are also evaluated in
Fourier space using the
convolution theorem. The nonlinear terms, Eqs
\eqref{coll1}-\eqref{coll2},
 are computed in real space. Note
that by using the convolution theorem we greatly speed up the
algorithm because the ballistic transport integrals are only
$\mathcal{O}(N\text{log}N)$ tasks, rather than $\mathcal{O}(N^2)$. For typical simulations this
means each time step is accomplished two orders of magnitude
faster than if conventional quadrature were utilised.

Viscous diffusion imposes the primary limitation on 
the time step $\Delta t$ in our problem. 
In order to avoid numerical instability, 
$\Delta t$ must satisfy a Courant condition,
\begin{align}
\Delta t < C\frac{(\Delta x)^2}{\text{max}[\mu]},
\end{align}
where $C$ is a constant, approximately $0.254$ 
for a third-order Runge-Kutta scheme on a
periodic domain (Canuto et al.~2006).  
We set $\Delta t$ in our simulations to the right-hand
side multiplied by a small `safety factor' $\sim 0.1$. 

The grid spacing $\Delta x$ is limited by the (physical) viscous
length $l_\mu$. The Gibbs phenomenon and worse ensues when the latter falls
beneath $\Delta x$, because unresolvable gradients develop which can
sharpen into discontinuities. 
In our units, the viscous length can be
approximated by $l_\mu
\approx \sqrt{\mu}$, and we take $\Delta x$ to be an order
of magnitude smaller. For additional safety in some runs we de-alias 
the solution, using the $2/3$ rule (Canuto et al.~2006).

\subsection{Boundary conditions} 

We apply two different boundary conditions to \eqref{GovEq}, either (a) periodic boundaries
or (b) `buffered' periodic boundaries. The former forces the
time-dependent solution $\tau(x,t)$ to satisfy
\begin{align}
\tau(0,t)= \tau(L,t).
\end{align}
Buffered boundaries, on the other hand, permit information to freely
leave the domain without re-entering it from the opposite
boundary. Periodic boundary conditions are retained, but we block wave
transmission at the boundaries by increasing $\mu$ in two buffer
zones encasing the two ends. Waves incident upon such zones
decay rapidly to zero before re-entering the domain on the other
side. 
A convenient
model profile for $\mu$ is
\begin{align}
\mu= \frac{0.2+\mu_0}{2}\left[\text{tanh}(x-L+l_B)-\text{tanh}(x-l_B) +2\right]+\mu_0,
\end{align}
where $l_B$ is the radial size of each buffer and $\mu_0$ is the
value of $\mu$ outside the buffers. This model lets
$\mu$ rapidly increase to 0.2 in the buffers, about twice the value
that can sustain BTI, as shown in Fig.~6 of Paper 1.

\subsection{Initial conditions and parameters}

The initial condition is usually small-amplitude white noise atop the
constant equilibrium state $\tau=\tau_0$, though
in certain simulations, such as in numerical tests and with B-ring
parameters,
we employ the exact nonlinear solutions computed in Paper 2. In the
simulations of ring edges, we set $\tau$ to a boxcar profile
\begin{align} 
\tau =
\begin{cases}
&1.5, \quad 30< x < 60 \\
&0.05, \quad \text{otherwise},
\end{cases} \label{edge}
\end{align}
and focus exclusively on the evolution of the inner edge near $x=30$. 
The simulation is terminated once
the inner and outer edge spread so far that they interact. 
When $L=100$ this occurs only at
extremely late times. The choice of upper and lower optical depths
(1.5 and 0.05 respectively)
ensures that the BTI fails to appear in either location for $\mu=0.02$
(see Fig.~6 in
Paper 1). 

The two main physical parameters in most runs are $\tau_0$ and the
parameter $\mu$. The former we set to either C-ring or inner B-ring values
$\tau_0\sim 0.1$ or $\tau_0 \sim 1$, and in almost all runs $\mu$ is
fixed at $0.025$, independent of $\tau$. The main numerical
parameters are $L$ and $\Delta x$. Because $l_\mu \approx 0.16$, we
set $\Delta x \approx 0.02$. The domain size $L$ we vary, but note
that for the local model to be a good approximation $L \ll r_0$. 
Our model space unit is $l_\text{th}$ which falls between
50 and 500 km (Paper 2). At $r_0\sim 10^5$ km, 
 it follows that $L$ should take values below 200
and 20 $l_\text{th}$ respectively.

\begin{figure}
\begin{center}
\scalebox{0.5}{\includegraphics{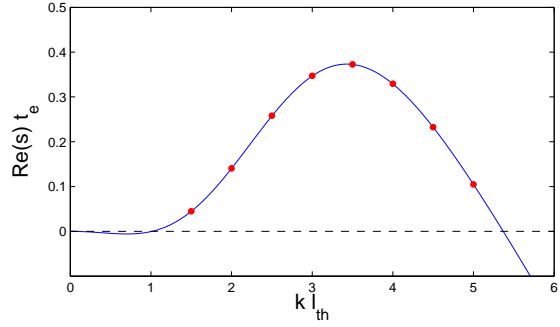}}
 \caption{Linear growth rates of the ballistic transport instability
   as a function of wavenumber
   $k$,
   when $\tau_0=0.5$ and
   $\mu=0.025$. The solid line represents the analytical theory (see
   Paper 1). The
   points are the growth rates computed from eight numerical
   simulations.}
\label{disp}
\end{center}
\end{figure}

\begin{figure}
\begin{center}
\scalebox{0.5}{\includegraphics{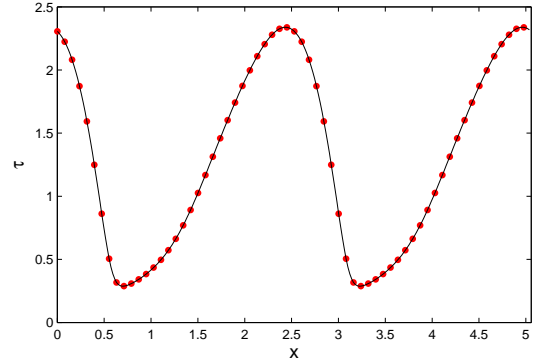}}
 \caption{Two periods of a nonlinear wavetrain calculated by the
   numerical simulation and the direct methods of Paper 2. The solid curve
   represents the latter, and the dots the former after being evolved
   forward by the code for 10 periods, i.e.\
   $t=2029$. Parameters are $\tau=1.3$,
    $\mu=0.025$, and the wavenumber is $k=2.4849$.
}\label{nlwtest}
\end{center}
\end{figure}

\subsection{Numerical tests}

We present two tests that demonstrate the accuracy of our numerical
tool. The first one checks that the code reproduces the linear growth
rates of low-amplitude disturbances. We seed two wavelengths of an
 unstable mode of
specified wavenumber $k$ and amplitude $\sim 10^{-6}$, and then 
evolve it forward until it grows by at
least two orders of magnitude. We subsequently measure the
growth rate and compare with the
analytic dispersion relation of the linear modes (cf.\ Eq.~(49) in
Paper 1). Our results are plotted in
Fig.~\ref{disp}. The agreement is good, with the relative errors below 1\%.

The second numerical test assesses the code's handling of nonlinear
waves. 
For initial conditions we employ
 two wavelengths of a nonlinear
solution to \eqref{GovEq} computed using the methods of
Paper 2 and thence propagate the wavetrain for exactly 10
periods. In Fig.~\ref{nlwtest} we show the result of a typical test;
the solid line represents the initial condition at $t=0$, and the dots
represent the wavetrain after exactly 10 periods, $t\approx 2029$. 
The excellent agreement in the shape and location of the
two waves confirm that the wave profiles and speed are accurately
reproduced by the code. Again, the relative error falls
below 1\%.

\section{C-ring simulations: low-amplitude wavetrains}

In this section we explore the low-$\tau$ regime relevant to the
C-ring. As discussed in Paper 2, the BTI in this setting
forms steady wavetrains of small amplitude, and we find
that it is the
emergence and competition between these structures that characterises
the nonlinear evolution of the instability.

\subsection{Periodic boundaries: free wavetrains}

Our initial run employs periodic boundary conditions, and sets
 $\tau_0=0.175$ with a constant $\mu=0.025$. The value for $\tau$ is
 slightly larger than observed but presents results that are easier to
 interpret. 
We restrict the
size of the domain $L$ to twenty times $l_\text{th}$. As argued in
Paper 2, structures in Saturn's C-ring indicate that $l_\text{th}\sim
500$ km, and so $20l_\text{th}$ approximately encompasses the extent
of wave activity in the C-ring. It is also near the local model's limit of
applicability. The initial condition is 
small-amplitude white noise $\sim 10^{-5}$, and the simulation is run for 1000 erosion
times.

\begin{figure}
\begin{center}
\scalebox{0.5}{\includegraphics{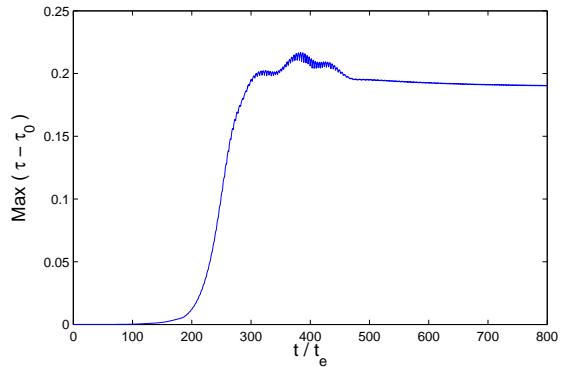}}
 \caption{Time evolution of the disturbance amplitude as a result of
   the BTI's growth (see Section 3.1). 
   The initial condition is small-amplitude white
   noise; physical parameters are $\tau_0=0.175$ and $\mu=0.025$,
   while numerical parameters are $L=20 l_\text{th}$ and $N=512$.}
 \label{hist1}
\end{center}
\end{figure}

\begin{figure*}
\begin{center}
\scalebox{1.0}{\includegraphics{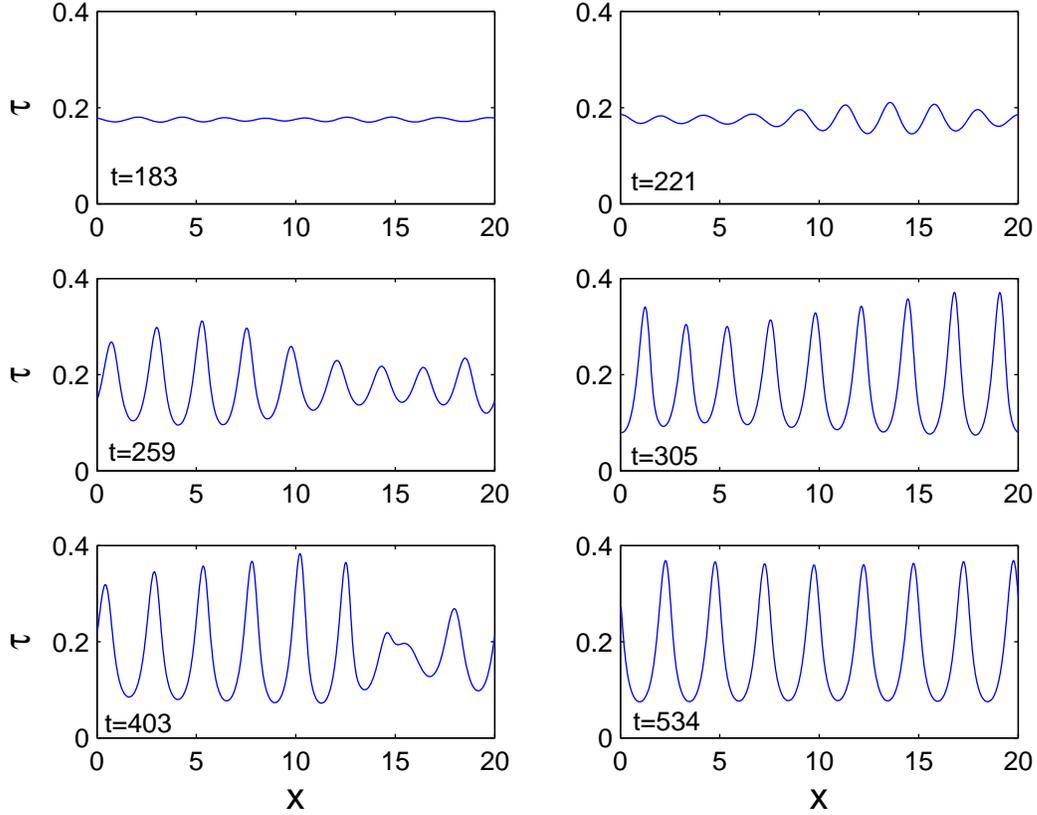}}
 \caption{Six snapshots of the evolution of $\tau$ due to the BTI, as
   described in Section 3.1. In
   all panels the wave crests are moving to the left (inward), though
   the longer waves in panel 6 do so extremely slowly.}
 \label{snaps1}
\end{center}
\end{figure*}

Results are plotted in Figs \ref{hist1} and \ref{snaps1}. The first
figure shows the evolution of the perturbation amplitude, which we
meaure by $\text{max}[\tau-\tau_0]$. The second
figure presents six snapshots of $\tau$ at different (unequally
spaced) times. The early 
stages of the evolution witness the independent and exponential growth
of competing linear BTI modes, with e-folding times $\sim 100$ (see
Fig.\ 1 in Paper 2). By $t=150$ the system is dominated by the fastest
growing mode, associated with the wavenumber $k=2.827$
(thus nine wavelengths fit into the domain). This mode, low in
amplitude and still
roughly sinusoidal in shape, can be observed in the first panel of
Fig.~\ref{snaps1}.

The exponential growth ceases after some 250
erosion times, as the system approaches the
 exact steady nonlinear wavetrain
solution associated with $k=2.827$. However, as panels 2-5 in
Fig.~\ref{snaps1} indicate, the waveform exhibits significant
modulational perturbations, conspicuous
 as early as $t=220$. 
These correspond to the action of a secondary instability upon the
$k=2.827$ wavetrain; as shown in Section 4 in Paper 2, the wavetrain
is too short to be stable for these parameters.
Not long after $t=400$ the secondary instability destroys the
$k=2.827$ solution, which is
superseded by the longer (and linearly stable) $k=2.513$ solution
(panel 6). Over the course of the next few hundred erosion
times, the amplitude slowly relaxes to the value predicted by Paper 2
(roughly 0.19, see Fig.~\ref{hist1}).    

This general pattern of behaviour is reproduced by all parameter
choices we have tried, and is thus not limited to the small $\tau_0$ regime. 
At first the fastest growing mode dominates the evolution
and the system approaches a wavetrain of the same wavenumber; but because
this wavetrain is always unstable, the system eventually migrates away
and seeks a linearly stable longer-wavelength solution. Depending on
the size of the domain (and hence the number of available modes), this
process can take one step (as above) or several steps. This is
in contrast to analogous behaviour in the viscous
overstability, where the wavelength selection procedure is more
involved --- mainly because the fastest growing wavelength and the
first stable wavelength are much further apart (Latter
\& Ogilvie 2009, 2010). Similar behaviour also occurs when the
 initial condition is a
 long unstable nonlinear wavetrain (with $k< 2.225$). 
 These
solutions break up relatively quickly and the system settles on a stable
shorter-wavelength wavetrain.

\begin{figure*}
\begin{center}
\scalebox{1.0}{\includegraphics{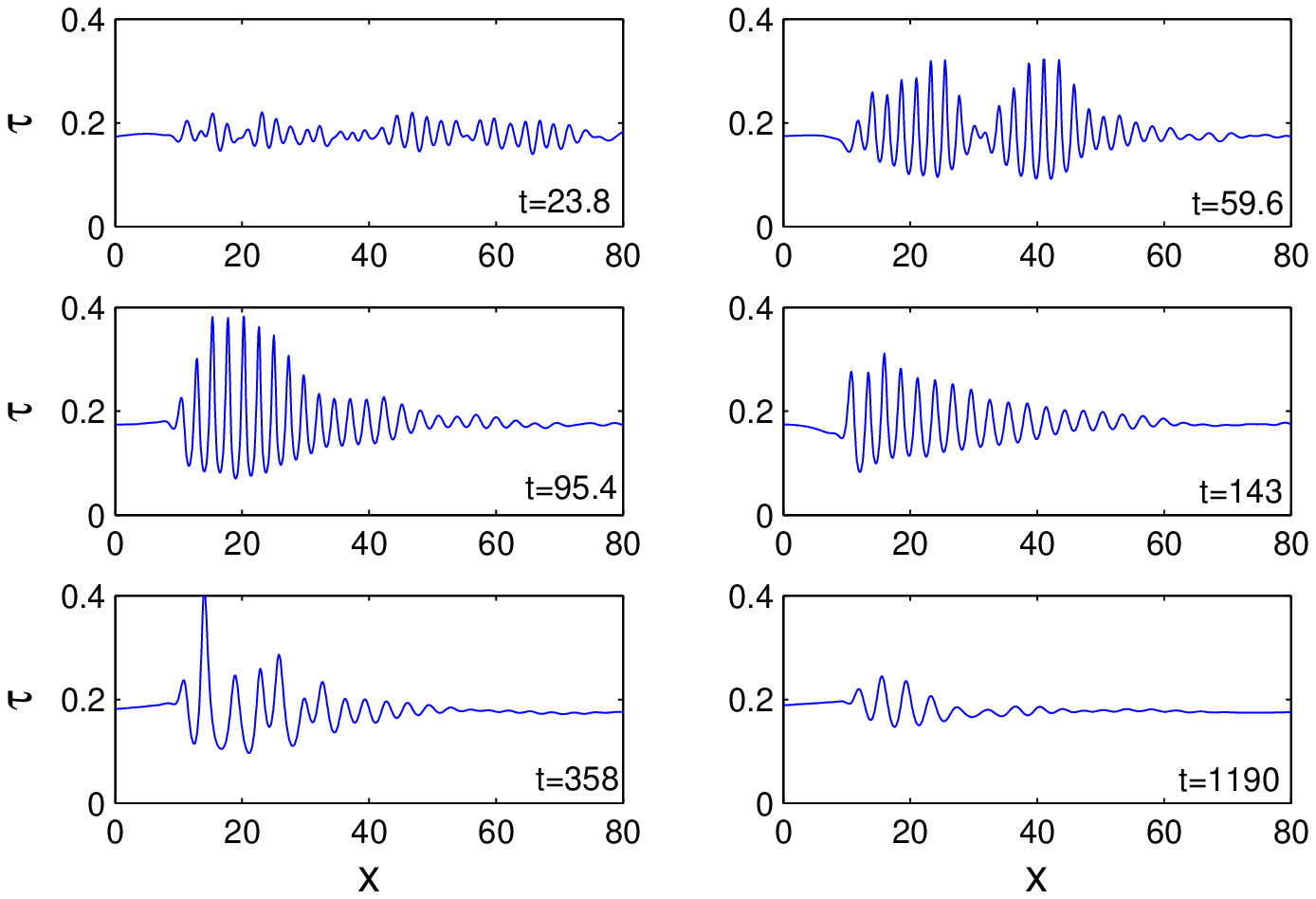}}
 \caption{Six snapshots of a run with buffered domains. Here $L=100$
   and $l_B=10$. Main parameters are $\tau=0.175$ and
   $\mu_0=0.025$. Waves crests propagate inward in panels 1 and 2, and mainly
 outward in panels 3-6. The group velocity of the disturbances, however, always
points inward.}\label{buff2}
\end{center}
\end{figure*}

\subsection{Buffered boundaries}

The periodic box simulations of Section 3.1
indicate that the BTI, when present, 
always takes the system to a uniform and stable travelling
wavetrain.
This outcome, however, could be viewed
 as an artefact of the boundary
conditions and the limited extent of the box.  
When periodic boundary conditions are imposed, the system
`senses' the translational symmetry of the domain after a sufficiently
long time and
is thus attracted to the steady wavetrain solutions admitted
by this symmetry.
But in the real rings there is no radial periodicity or global
 translational symmetry.
As a consequence, steady uniform wavetrains
are not exact nonlinear solutions globally -- though they are approximate
solutions locally -- and hence cannot function as global attractors. 
The real rings will certainly exhibit nonlinear wavetrains, and the
secondary instabilities that assail them, yet the larger-scale dynamics
may be rather different to that shown in Section 3.1. They may instead
exhibit a competition between different
wavetrain solutions, propagating towards or away from each other,
partly controlled by underlying gradients and inhomogeneities in the
background state. 

This subsection probes some of this
behaviour by breaking the translational symmetry of the box. Buffer
regions are imposed across both boundaries, as decribed in Section
2.3. These eliminate the attracting stable wavetrain solutions from
the phase space, thus permitting a more realistic and interesting set
of dynamics to develop. 
Our main simulation takes a mean optical depth 
$\tau_0=0.175$, a large domain $L=100$, and buffers of size $l_B=10$.
White noise of moderate amplitude $(\sim 0.01)$ 
is seeded in order to speed up
the evolution. The resulting evolution is summarised in 
Fig.~\ref{buff2}, where we show six snapshots.

The fastest growing modes dominate the initial stages of the
evolution, and after growing to appreciable amplitudes  
they undergo complicated interactions involving beating patterns and
the formation of localised wave packets (panels 1 and 2). 
The group velocity of these
waves $c_g$ is negative and relatively large $\sim -0.5$ (see Fig.~3b in
Paper 2). As a result,
disturbances propagate to the inner boundary on relatively short
timescales
where, over the course of the simulation, they mostly disappear into the buffer
zone. Ultimately the system settles into a quasi-steady state
characterised by low-amplitude long-wavelength waves with positive phase
speeds (panel 6). These waves emerge from a `source' at the inner
buffer and decay to zero as their crests propagate outwards. Though
their phase speed $c_p$ is positive, the group velocity
of these long waves remains negative (Fig.~3b, Paper 2); 
thus disturbances at larger $x$ cannot be replenished by activity
in the inner regions. 
Activity persists in the inner regions, on the other hand, as
`residue' of the nonlinear behaviour in the first 100 erosion times. 
    
The key feature controlling these dynamics is the `convective', as
opposed to `absolute', nature of the BTI: unstable waves do not
grow in place, but travel as they grow. A disturbance will reach
large amplitudes but it may leave the region of interest before it
does so, especially if the growth time (the inverse of the linear
growth rate $s$) is longer or similar to   
$L/c_g$, i.e. the time it takes for information to traverse the
domain. As a consequence, any given ring region may eventually return
to its undisturbed state. (This scenario is in marked contrast 
to the unbuffered periodic simulation of
Section 3.1, where travelling disturbances are never lost and indeed can
positively reinforce themselves.)
In the buffered simulation described by Fig.~\ref{buff2}, 
$1/s \sim L/c_g$.
As a result, the disturbances just attain sufficiently large amplitudes for
nonlinear interactions to ensue, and hence for activity to be
sustained, though at a fairly low level.

\begin{figure}
\begin{center}
\scalebox{0.55}{\includegraphics{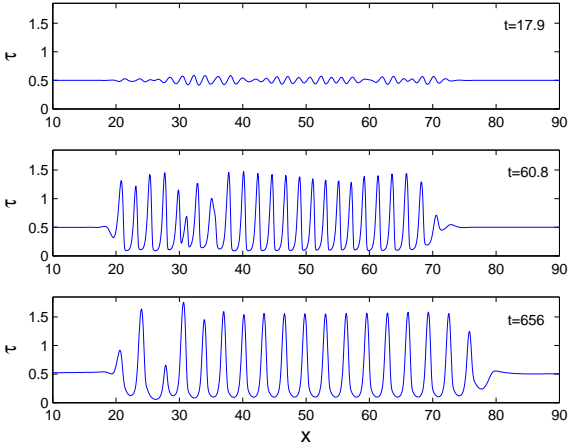}}
 \caption{Three snapshots of a run with buffered boundaries. Here
   $L=100$ and $l_B=20$. The mean optical depth is $\tau=0.5$, and
   $\mu_0=0.025$. Waves characterise the simulation, the crests propagating
   inward in the first two panels and outward in the last. }\label{buff1}
\end{center}
\end{figure}

To further test this idea, we conducted additional simulations, one with $L$
reduced by half and another with smaller amplitude initial
conditions $\sim 10^{-4}$. In both simulations, growing disturbances propagate out
of the active zone before they achieve significant amplitudes, 
and eventually the
ring returns to a near homogeneous state. 
We also ran a simulation
illustrating the opposite extreme, in
which the growth time $1/s$ is much smaller than $L/c_p$; we set
$\tau$ to 0.5 and hence $s\sim 0.4$, an order of magnitude larger than
in Fig.~5. The results of this
last simulation are summarised in Fig.~\ref{buff1}, which displays three
snapshots at different times. Note that here $l_B=20$.         
Initially, the fastest-growing modes 
dominate the active region, leading to the emergence
of wave packets travelling inwards.
After an internal wavelength selection process (panel 2), 
the system settles on a long wavetrain with
$ k\approx 1.96$ (panel 3) and positive $c_p$.
 The key point is that, because of the large growth rates,
disturbances attain large amplitudes before hitting the inner
buffer, allowing the nonlinear dynamics sufficient time to develop 
and sustain activity throughout the \emph{entire} central region. 

 We expect some of this behaviour to characterise
the low-amplitude undulations in the C-ring, which occur in a regime
near marginal stability and in a relatively small domain ($\sim 20
l_\text{th}$).
If left
to evolve independently for sufficient time, it is possible that the
present undulations may simply propagate out of the region and disappear at
the inner edge of the C-ring. 
Alternatively, larger amplitude disturbances may have already
left the system, leaving behind the low-amplitude waves we see today;
these may then correspond to the undulations in panel 6 in Fig.~\ref{buff2}.
A third, more likely scenario, is that the C-ring suffers a
a continual supply of noise which reseeds the BTI and replenishes the
undulations. A complication in all three hypotheses is that the waves   
travel through an
environment punctuated by disruptive features, such as plateaus,
ringlets, and gaps, as well as possible large-scale gradients in ring
properties (Figs 13.17 and 13.21 in Colwell et al.~2009, Hedman et al.~2013, Fillachione et
al.~2013).
Finally, we note
the longer wavelengths produced by our simulations ($k\approx 1.5$), which 
suggest an estimate for $l_\text{th}$ closer to 300 km, shorter than
estimates based on linear stability (see Section 4 in paper 2).

\section{Inner B-ring simulations: wavetrain pulses}

This section presents simulations that approximate conditions in the
inner B-ring. As discussed in Paper 2, this region is of special
interest because it can exhibit bistability, whereby the system
falls into a
homogeneous `flat' state, or a small set of large-amplitude `wave'
states. It is then possible that
the ring radially breaks up into
adjoining flat and wave zones, with the boundaries between the regions
undergoing their own dynamics. We focus on this scenario
here and show that this is indeed possible. 

Our fiducial B-ring run takes parameters $\tau_0=1.3$ and $\mu=0.025$, 
which admit the linearly
stable homogeneous state $\tau=\tau_0$. As explored in Section 3.3 in 
Paper 2, these parameters
also support a number of travelling nonlinear wavetrains.
Our initial condition comprises a portion of a wavetrain, possessing
$k=2.5$, inserted between $x=10$ and $x=30$. The remainder of
the domain is set to $\tau=\tau_0$. The two ends of the wavetrain are
abruptly smoothed to the homogeneous value over a few grid cells.
The initial condition is hence a `wave packet' or `wave zone'.
The size of the box is $L=50$, and we reinstate the periodic boundary
conditions (without buffers).

\begin{figure*}
\begin{center}
\scalebox{0.85}{\includegraphics{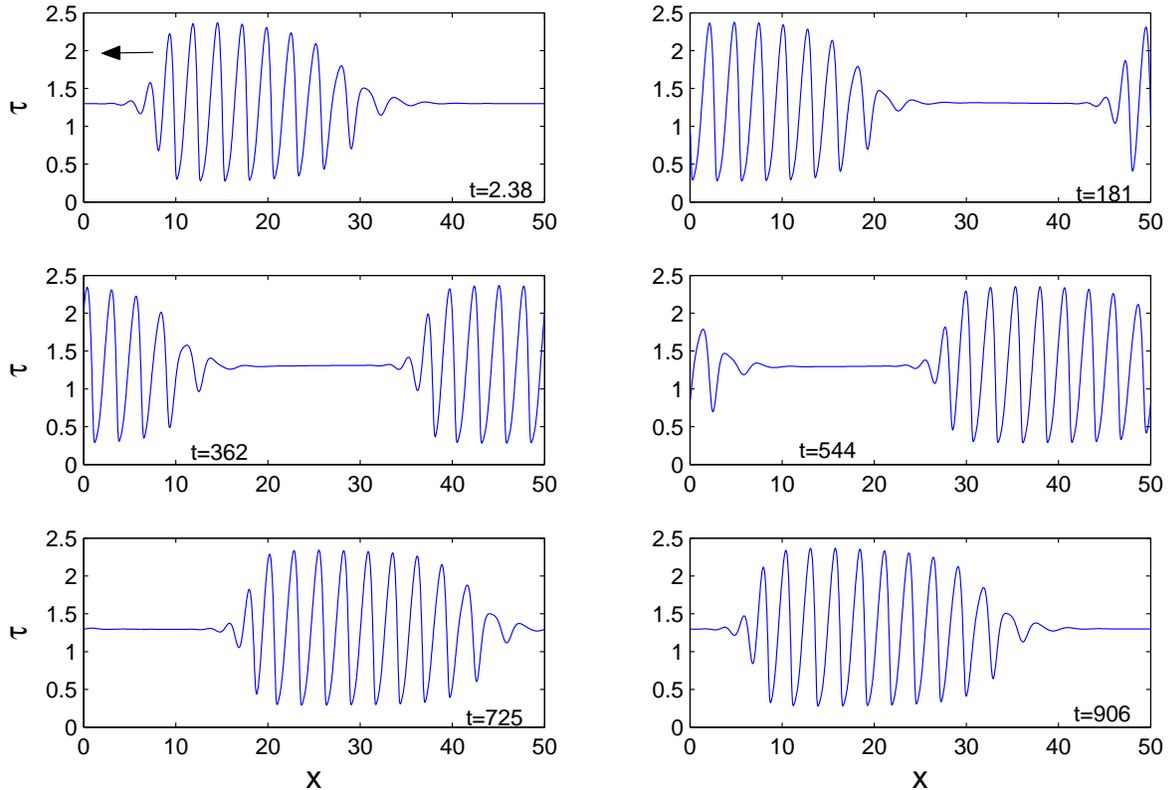}}
 \caption{ Six snapshots showing the evolution of BTI wavepacket
   propagating through a bistable ring. The background is described by
   $\tau_0=1.3$, $\mu=0.025$, and the numerical parameters are
$L=50 l_\text{th}$, and $N=1024$. The wave packet and its constituent
waves travel
to the left at different speeds. In addition, the front of the packet
travels faster than the rear of the packet, leading to its
spreading. An arrow is included indicating the direction of
propagation.}
\label{snaps2}
\end{center}
\end{figure*}

Six snapshots of the evolution are displayed in Figure \ref{snaps2} at
equally spaced times. We find that within the packet the individual peaks of
the wavetrain travel to the left with the phase speed $c_p\approx
-0.014 $ predicted by the
calculations of Paper 2. Moreover, the wavepacket as a whole 
travels to the left with the faster group speed $c_g=d\omega/dk\approx -0.058$,
associated with $k=2.5$. As the wave
packet propagates it also spreads, which is clear from the first and
last snapshots in Fig.~\ref{snaps2}: after one traversal of the domain
the rear of the packet has lagged behind the packet's leading front. 
This is because the packet's
component wavelengths at its front are shorter than at its rear.
Consequently
$c_g$ decreases throughout the packet. 
The rear moves slower, and the structure as
a whole spreads.

The wave packets, understood as
homoclinic orbits in phase space, share a number of features with
localised oscillatory states in a variety of driven and/or
bistable
systems. Notable examples include
 the complex Ginzburg Landau equation (Aranson
\& Kramer 2002, Burke et al.~2008), plane Couette flow (Schneider et
al.~2010), 
and magnetoconvection (Buckley \&
Bushby 2013). 
However, because $c_g$ varies throughout the packet and $\neq c_p$, 
our BTI structures
  resist
the usual techniques of dynamical systems (e.g.\
Burke \& Knobloch 2007). In particular, as the packet
spreads in time, there are no distinct states
comprising the classical `snakes and ladders' pattern in the phase
space. Instead, over time, the system progresses through a continuum
of such states.

Qualitatively, these results agree relatively well with
Cassini observations which reveal that the inner B-ring radially 
splits into two `flat' zones, where the photometric $\tau$ is
constant, circumscribed by three `wave' zones, exhibiting nonlinear
wavetrains of wavelength $\sim 100$ km and amplitude $\sim 1$ in
$\tau$ (Figs 13.11 and 13.13 in Colwell et al.~2009). We hence identify the flat zones with
the inactive homogeneous state $\tau=\tau_0$ and the wave zones with
travelling (and spreading) wave packets. 
Our results then suggest that the two flat zones are shrinking and will
 `evaporate' entirely in $\sim 100 t_e$. Of course, the
Janus/Epimetheus 2:1 inner Lindblad resonance situated between the two
flat zones does complicate things, but the overall agreement
encourages us to view the inner B-ring
 as controlled by the BTI's bistable dynamics.

A number of intriguing discrepancies remain, however. 
As discussed in Paper 2, the troughs of the theoretical wave profiles
(measured in dynamical $\tau$)
 are lower
 than those exhibited by the observed waves (measured in photometric
 $\tau$). 
The disagreement could disappear once the differences between photometric and
 dynamical optical depth are corrected for, but this is not yet certain.
Another, potentially related, problem issues from the discrepancy 
between the observed \emph{mean} photometric
 optical depth in the flat and wave zones (Colwell et al.~2009).
 If this indeed corresponds
 to significant differences in the surface density (hence dynamical
 $\tau$), 
 then our
 picture of the B-ring dynamics will need to be modified. A final point is 
that it can be difficult to seed a travelling wave packet in our
theoretical simulations; an arbitrary
 localised large-amplitude perturbation is
usually insufficient. A wavelike or very large-amplitude disturbance
does better. Sources for appropriate disturbances may be present in
the Janus/Epimetheus resonace, the inner B-ring edge, or the
transition to the high-$\tau$ regions in the mid B-ring, but again
this is uncertain.

\section{Simulations of the inner B-ring edge}

We finish our numerical study with a handful of exploratory simulations of 
a sharp ring edge spreading under the action of viscosity and ballistic transport.
So far we have reproduced the dynamics of the BTI mostly in isolation
of 
 strong inhomogeneities. In this section 
we see how it fares in the presence of a dramatic variation in $\tau$.

We employ an initial condition representing an isolated wide ringlet possessing
extremely sharp edges, as described by
Eq.~\eqref{edge} in Section 2.4. In order to simplify the numerical results,
we take the optical depth of the ringlet to be sufficiently high so that no wavetrains are possible, thus $\tau=1.5$.
Similarly, in the surrounding medium we take $\tau$ to be sufficiently low so that the BTI is suppressed.
We then focus exclusively on the dynamics of the \emph{inner edge} of the ringlet, 
and treat it as an approximation to the inner edge of the B-ring. 
The viscosity, represented by the parameter $\mu$, is expected to vary between the 
low and high $\tau$ regions in our setup (Araki \& Tremaine 1986,
Wisdom \& Tremaine 1988, Daisaka et al.~2001). 
However, for our main runs we let it remain a constant; additional simulations 
in which $\mu$ is a power law in $\tau$ do not exhibit qualitatively
different results.
In keeping with previous work, we associate $t_e$ with
the erosion time of the high-$\tau$ region.

\begin{figure}
\begin{center}
\scalebox{0.55}{\includegraphics{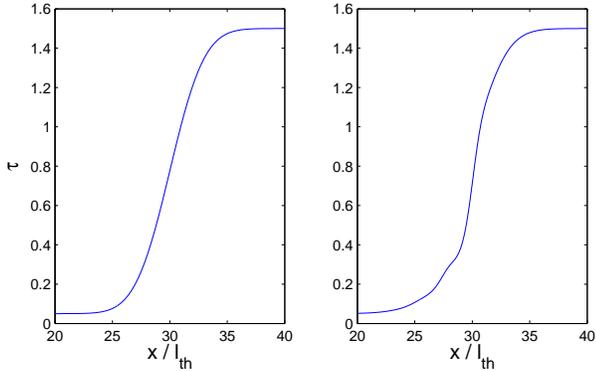}}
 \caption{The $\tau$ profile of two ring edges at $t=43.56$. The
   initial condition in both cases is a step function, but the left
   panel shows an edge evolved under the action of viscous diffusion alone, 
   while the right shows spreading
   under the action of both ballistic and viscous transport. 
   In both simulations $\mu=0.065$ and is constant. }\label{edge1}
\end{center}
\end{figure}

\begin{figure*}
\begin{center}
\scalebox{0.8}{\includegraphics{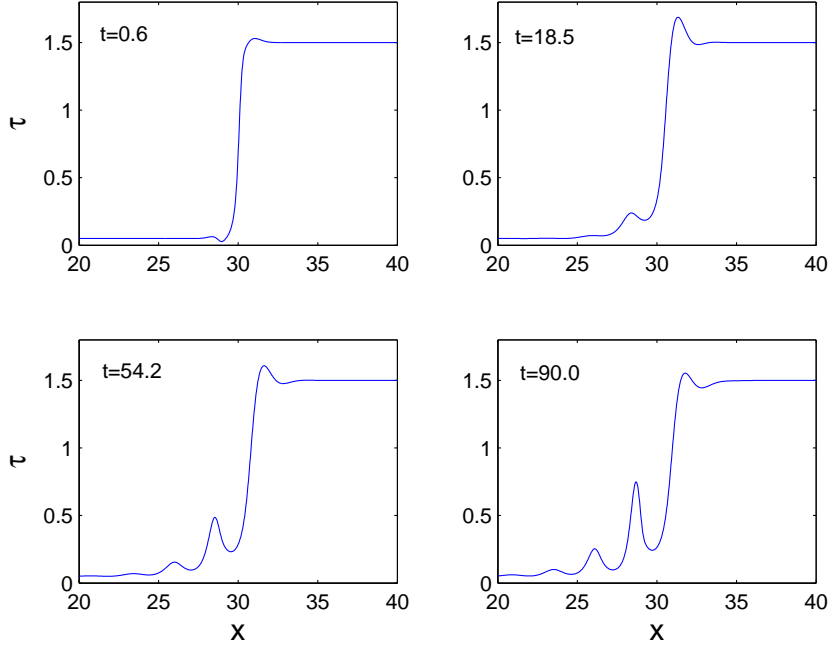}}
 \caption{Four snapshots showing the spreading of a sharp edge with
   $\mu=0.025$ and constant.}\label{edge2}
\end{center}
\end{figure*}

Our first runs take a relatively large $\mu$, equal to
$0.65$. Consequently, the BTI will not work for any $\tau$
(see Fig.~2 in Paper 2). This setup permits us to study the influence of
ballistic transport on ring spreading in isolation of the instability
itself. 
For comparison, we also perform a simulation with ballistic transport
completely de-activated, leaving the edge to spread via viscous
diffusion alone.
Figure \ref{edge1} shows two snapshots taken at 
the same time from these two simulations.
While the purely viscous simulation exhibits
the self-similar tanh solution of the classical diffusion equation, the BT simulation reveals a more complex structure. 
A characteristic `ramp' profile develops at the foot of the edge, exhibiting a shallower gradient than the tanh
profile. Meanwhile the edge itself remains sharper in comparison to pure viscous spreading. 

These results should be compared with the
more viscous runs of D92; see their Figs 4 and 5 (for which $Y=3\times 10^4$ and $Y= 1\times 10^5$ respectively). 
The most viscous D92 run (their Fig.~4) resembles our
Fig.~\ref{edge1}a, presumably because BT is sub-dominant. The corresponding $\mu$ is
difficult to calculate because of uncertainty in the value of $l_\text{th}$, and also because the D92 viscosity is not
constant. Its mean value we estimate to be $\gtrsim 0.1$, consistent with our results. D92's intermediate run (their
Fig.~5), yields an edge profile similar to our Fig.~\ref{edge1}b, and indeed possesses a comparable mean $\mu$ (roughly 0.05).
The main difference is the absence of a `hump' on the high-$\tau$ side of the edge. In fact, our $\mu=0.065$ simulation 
does exhibit a hump at earlier times, but it has diffused away by
$t=10$. Perhaps this small discrepancy is due to our adoption
of the simpler absorption probability function $P$ (see discussion in
Papers 1 and 2). Overall, however, we confirm the findings of D92 that ballistic transport can
maintain the sharpness of a spreading edge, and concurrently develop a `ramp-like' feature at its foot.

The ramp in Figure 8b possesses values of $\tau$ which place it in the favourable range for BTI (see Fig.~2 in Paper 2).
However, our choice of $\mu$ prevents instability. To test whether the ramp region can indeed support
instability, we reduce $\mu$ to a value of $0.025$ and redo the numerical calculation. In Fig.~\ref{edge2} we present
six snapshots of the ensuing evolution. Indeed, as early as $t=18$ a growing undulation appears at the foot of the edge
(panel 2). As the ramp expands and generates greater $\tau$, more and more of the region becomes unstable
leading to the emergence of further growing modes. Throughout this phase, the edge itself remains relatively sharp and now exhibits
the hump on the high $\tau$ side. At later times the ensuing wavetrain amplitudes reach levels $\sim 1$, and the waves
propagate slowly inwards. 

Similar growing modes are not observed in D92's low viscosity run (see
their Fig.~6, which has $Y=3\times 10^5$), even though it possesses a
comparable mean $\mu$ ($\sim 0.01$). 
The most prominent feature in the D92 run, in fact, is the high $\tau$ wavetrain, which we
have suppressed by our choice of $\tau$ in the optically thick
ringlet. 
Note, however, that the D92 simulation is only run till $t=41$, and so it is
possible that appreciable BTI waves would have emerged at later
times. Moreover, more recent runs by the same group witness growing
`humps' in the ramp for certain parameter choices (Durisen, private
communication). We conclude that ramp stability is sensitive to input
physics and its parameters.
 
Observations of the inner B-ring edge do not indicate the existence of BTI; the ramp, in fact, exhibits a remarkably
unblemished linear profile (Fig.~13.21, Colwell et al.~2009). It is also unlikely that the C-ring plateaus are related to the
emergence of these BTI undulations. The morphologies of the two are dissimilar and
there is no positive gradient in background $\tau$ where plateaus
exist, in contrast to our simulations; the observations indicate that
ramps and plateaus appear separately.
 We hence conclude that the BTI is suppressed at the inner B-ring edge, though the responsible
physical process is unclear, and obviously absent from our model. 
It is possible that our choice of distribution function $f$, or the parameters
that appear in it, may unrealistically boost instability in this
region. 
But more work, with further
refinements, is needed before this can be verified.

Finally, we speculate that, because the ramp increases in size as the
edge spreads, the ramp width could be used as a diagnostic
for the ring spreading time, possibly even allowing researchers to
construct past morphologies and probe ring formation scenarios. For
example, the
observed ramp is $\approx 1000$ km, or possibly $\approx
10l_\text{th}$; our simulations indicate that such a structure
would take $\approx 100 t_e$ to form, if the edge was initially very
sharp.

\section{Conclusion}

In this paper we have developed a reliable and efficient numerical
tool with which to simulate the ballistic transport process in
planetary rings. We have subsequently reproduced the nonlinear
evolution of the BTI in models of both the C-ring and B-ring, in
addition to the spreading of a ring edge. 

Both our B-ring and C-ring
simulations
validate the predictions of Paper 2's semi-analytic theory.  
C-ring simulations saturate by forming stable low-amplitude wavetrains within a
narrow range of preferred wavelengths. But because of the translational
symmetry of our local model this outcome is not a global
solution to the real, radially structured, C-ring. In order to eliminate
some of the unrealistic effects of the translational symmetry,
additional simulations were conducted
with buffered boundaries.
 Near
marginality, as in the C-ring, BTI modes
possesses low growth rates and yet retain
relatively large group velocities and,
as a consequence, the BTI's `convective' character
becomes important. 
Our buffered simulations show that unstable disturbances can propagate out of the
region of interest before their nonlinear dynamics develop and
sustain appreciable amplitudes. Unless continually fed new perturbations,
the BTI may saturate at a very low level of activity.
 It is possible that the C-ring
has fallen into such a state.
However, C-ring undulations
 compete with other features, such as plateaus, ringlets, and gaps
that may interfere with their evolution, or alternatively help seed
fresh BTI modes. Undoubtedly the dynamics are complicated in this
region and more irregular than predicted by the local shearing
sheet with pure periodic boundary conditions.

Our simulations of the inner B-ring show that it is possible that the system
splits into stable wave zones and stable homogeneous zones, in
agreement with Cassini data.
Moreover, the wave zones propagate through the homogeneous regions as
independent wave packets 
while simultaneously spreading. Consequently, both the observed larger and
smaller flat spots may evaporate in a time $\sim 100t_e \sim
10^7-10^9$ yr. However, as discussed in Paper 2, the morphologies of the
theoretical and observed B-ring
waves exhibit troubling discrepancies. The troughs of the former are
too deep, while the mean $\tau$ in the former varies between flat and
wavy zones. The disagreement could issue from the simple fact that the
theoretical and observational profiles are measured in dynamical and
photometric optical depths, respectively. But further work is needed
to establish this directly.  

Finally, we conducted exploratory simulations of the inner B-ring
edge. We find that ballistic transport does not arrest the viscous 
spreading of the edge, but resculpts it as it spreads. In
particular, ballistic transport forms a ramp-like feature at its base, 
while maintaining the sharpness of the edge --- in agreement
with previous simulations. We also find, in low viscosity runs, that
the ramp structure is susceptible to BTI. But as the observed ramp 
does not exhibit wave (or any other) features, we conclude that the
BTI is suppressed in this region by physics not captured in our
fiducial simulations.

In the future we hope to refine our physical model 
and conduct
more detailed simulations, especially of spreading edges. 
There are two obvious
improvements: 
the inclusion of a $\tau$-dependent viscosity, and an absorption probability $P$ that 
depends on $\tau$ at both the emitting
and absorbing radius. 
Preliminary simulations indicate little qualitative
change when $\mu$ is an increasing function of
$\tau$. On the other hand, a better $P$ model may alter 
our results more significantly, not only for the
C-ring dynamics but for conditions at the inner B-ring edge. Indeed,
it may aid in the suppresion of the BTI in the
ramp region. Unfortunately the more realistic $P$ precludes
use of the convolution theorem and its many computational
benefits. An intermediate model, however, need only incorporate the first few
terms in an expansion of $P$ (i.e.\ the first few ejecta ring-plane crossings),
and the convolution theorem could then be applied to each term. 
 
Simulations with the improved model may better reproduce, and help
explain, the observed morphologies of spreading edges.
They may also constrain the spreading
time of the inner A-ring and B-ring by examining the widths of their
ramp regions. 
Other targets for research include the C-ring plateaus. BTI does not
generate these features and, being too narrow, nor does it emerge
within them. But ballistic transport may sculpt
their structure, and in particular sharpen their front and rear
edges. Future numerical work here would complement previous
investigations which used a descendent of the D92 code (Estrada \&
Durisen 2010). It could also explore a possible connection between the C-ring
plateaus and similarly sized narrow rings, such as the $\epsilon$ ring
in the Uranian system.

\section*{Acknowledgments}

The authors would like to thank the reviewer, Dick Durisen, for a
thorough and helpful review that improved the quality of the manuscript.
This research was supported by STFC grants ST/G002584/1 and ST/J001570/1.

\end{document}